\documentclass[aps
,rmp
,twocolumn%
,amsmath
,amssymbol
,showpacs%
,superscriptaddress,floats%
]{revtex4}

\usepackage{graphicx}
\usepackage{times}

\newcommand{\be}{\begin{equation}}
\newcommand{\ee}{\end{equation}}
\newcommand{\bea}{\begin{eqnarray}}
\newcommand{\eea}{\end{eqnarray}}
\newcommand{\kT}{k_{\rm B}T}
\newcommand{\e}{{\rm e}}
\newcommand{\dd}{{\rm d}}

\newcommand{\affila}{ 
 Dept.\ of Biological Physics, E\"otv\"os University,
 P\'azm\'any P.\ stny.\ 1A, H-1117 Budapest, Hungary
}
\newcommand{\affilb}{
 Laboratoire de Physico-Chimie Th\'eorique, UMR 7083 CNRS, ESPCI,
 10 rue Vauquelin, F-75231 Paris C\'edex 05, France
}

\begin{document}

\title{
Effects of intermediate bound states in dynamic force spectroscopy
}

\author{Imre Der\'enyi}
\email{Derenyi@elte.hu}
\affiliation{\affila}

\author{Denis Bartolo}
\email{Denis.Bartolo@espci.fr}
\affiliation{\affilb}

\author{Armand Ajdari}
\email{armand@turner.pct.espci.fr}
\affiliation{\affilb}

%\date[]{\protect\today}
%\date[]{submitted to where..., when...}

\begin{abstract}
We revisit here some aspects of the interpretation of dynamic force spectroscopy experiments. The standard theory predicts 
a typical unbinding force $f^*$ linearly proportional to the
logarithm of the loading rate $r$ when a single energetical barrier
controls the unbinding process; for a more complex situation of $N$
barriers, it predicts at most $N$ linear segments for
the $f^*$ vs.\ $\log(r)$ curve, each segment characterizing a different
barrier. We here extend this existing picture using a refined approximation,
we provide a more general analytical formula,
and show that in principle up to $N(N+1)/2$ segments can show up experimentally. As a consequence the interpretation of data can be ambiguous,
for the characteristics and even the number of barriers. A further 
possible outcome of a
multiple-barrier landscape is a bimodal or multimodal
distribution of the unbinding force at a given loading rate, a feature
recently
observed experimentally.
\end{abstract}

% 02.50.-r   Probability theory, stochastic processes, and statistics
% 05.40.-a   Fluctuation phenomena, random processes, noise, and Brownian motion
% 33.15.-e   Properties of molecules
%   33.15.Fm   Bond strengths, dissociation energies
% 68.43.-h   Chemisorption/physisorption:  adsorbates on surfaces
%   68.43.Mn  Adsorption/desorption kinetics
% 82.20.-w Chemical kinetics and dynamics
%   82.20.Db Transition state theory and statistical theories of rate constants
%   82.20.Kh Potential energy surfaces for chemical reactions
% 82.37.-j Single molecule kinetics
%   82.37.Gk STM and AFM manipulations of a single molecule
%   82.37.Np Single molecule reaction kinetics, dissociation, etc.
%   82.37.Rs Single molecule manipulation of proteins and other biological molecules
% 87.15.-v   Biomolecules: structure and physical properties
%   87.15.By   Structure and bonding
%   87.15.He   Dynamics and conformational changes
%   87.15.La    Mechanical properties
% 87.64.-t   Spectroscopic and microscopic techniques in biophysics and medical physics

\maketitle
\section{Introduction}
The last decades have witnessed a 
revolution in the methods to observe and manipulate single bio-macromolecules or bio-molecules complexes. New micromanipulation techniques have especially been put forward to probe the folded structure of proteins and to quantify the strength of adhesion complexes~\cite{titin,Poi01,Sim99,Nis00,Pie96}. An important 
 step in this direction
is the proposal of the group of Evans
to use soft structures to pull on adhesion complexes
or molecules at various loading rates (dynamic force spectroscopy or DFS)~\cite{Eva}.
Moving the other end of the soft structure at constant velocity
induces on the complex a pulling force that
increases linearly in time $f=rt$.
Measuring the typical unbinding time $t^*$
yields an unbinding force $f^*=rt^*$ that depends
on the pulling rate $r$. The typical outcome of such experiments is a plot of $f^*$ vs.\ $\log r$ composed by a succession of straight lines with increasing slopes (force spectrum). 
It has been argued and that it is possible to deduce the value of some relevant structural parameters of the complex by analysing force spectra thanks to an adiabatic Kramers picture. This picture consist in considering the unbinding process as the thermally activated escape from bound states over a succession of barriers along a one-dimensional path crossing a mountainous energy landscape~\cite{Eva}. Within this scheme, each straight line of the force spectrum witnesses the overcome of an energy barrier and its slope maps the barrier to a distance $x$ along the pulling direction. This procedure has been shown to yield reasonable values for a few systems, and has been conforted
by numerical simulations \cite{Hey00}.

Subsequently, theoretical studies have refined the above original model, e.g., by inclusion of rebinding events~\cite{Sei02}, study of time dependent loading rates~\cite{Eva}, or consideration of more complex 
topographies ~\cite{Str00} and  topologies of the energetical landscape ~\cite{Bar02}. 

In the present paper we explore the potential influence of the 
existence of intermediate bound states on the experimental dynamic response of adhesion complexes as probed in DFS. To achieve this goal, we first revisit the analysis of the escape from a bound state consisting of an arbitrary number of barriers along a 1D path under the application of an external load
(in line with earlier studies of Strunz et al. \cite{Str00}),
and then discuss the implications of this analysis for the interpretation of experimental data. 
In (\ref{sec_std}) the standard picture is recalled, together with its two underlying assumptions. In (IV), we first relax the {\it a priori} assumption of a deep fundamental bound state and provide a general expression that relates the typical rupture force to the loading rate (within a single escape rate approximation).
The practical implications of this new formula [Eq.\ \eqref{eq_fr}] are discussed, and in particular we comment upon intrinsic ambiguities 
in inferring informnation from a $[\log(r),f^*]$ plot. 
Then, we show in (V) that in the presence of multiple bound states 
it may be necessary to relax the other assumption
(a single typical rupture force for each loading rate)
as multimodal rupture force distributions naturally show up,
a feature recently observed in lipid extraction experiments~\cite{Eva02}.
\section{Model and Notations} 
\begin{figure}[!t]
\centerline{\includegraphics[width=.8\columnwidth]{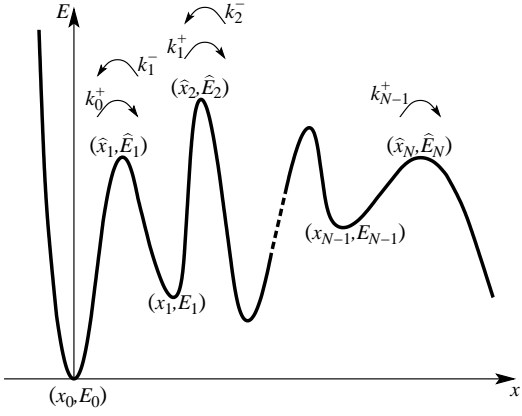}}
\caption{
Sketch of the one-dimensional energy landscape describing the unbinding pathway projected along the pulling direction. %
%The location $x_0$ and energy $E_0$
%of the fundamental bound state are arbitrarily taken as origins of the spatial and %energetical axes. 
}\label{fig1}\end{figure}
Figure \ref{fig1} illustrates the energy landscape of a one-dimensional
escape path with $N$ energy barriers and wells. The position and energy
of the $i$th energy well ($i=0$ marking the fundamental bound state,
and $1\leq i \leq N-1$ the intermediate ones) are denoted by $x_i$ and
$E_i$, respectively. Similarly, the position and energy of the $j$th
energy barrier are denoted by $\widehat{x}_j$ and $\widehat{E}_j$,
respectively (where $1\leq j \leq N$). For convenience, without loosing
generality, we set $x_{0}=0$ and $E_{0}=0$ for the fundamental bound
state. The unbound ``state'' is on the right hand side of the $N$th
barrier. If the energy differences $(\widehat{E}_{i}-E_i)$ and $(\widehat{E}_{i+1}-E_i)$ exceed $k_{\rm B}T$ the transition rates $k_i^-$ (and $k_i^+$) from the $i$th energy well over the left $i$th barrier (and right $(i+1)$th barrier, respectively) can be written according to the Kramers formula   
\bea
k^-_i &=& \omega_0 \, \alpha_i \, \widehat{\alpha}_i \,
 \e^{-(\widehat{E}_i-E_i)/\kT} ,
 %\qquad \mbox{and}
\label{eq_k-}\\
k^+_i &=& \omega_0 \, \alpha_i \, \widehat{\alpha}_{i+1} \,
 \e^{-(\widehat{E}_{i+1}-E_i)/\kT} ,
\label{eq_k+}
\eea
where
$\omega_0$ is a typical attempt frequency,
$\alpha_i$ and $\widehat{\alpha}_j$ are geometric factors
characterizing the shape of the $i$th energy well and $j$th energy
barrier, respectively. Note that there is no transition from the
fundamental bound state to the left, therefore, $k^-_0\equiv 0$.

We assume throughout the paper that the energy wells and barriers are sharp,
so that for any loading force $f$ their locations remain constant,
and their energies change as
$E_i(f)=E_i(0)-fx_i$ and 
$\widehat{E}_j(f)=\widehat{E}_j(0)-f\widehat{x}_j$.
To simplify the notations, wherever the argument of the energies and
transition rates is omitted, a loading force $f$ is implicitly
assumed. 

Finally it will prove convenient to introduce a few compact notations.
For any $0\leq i<j\leq N$ we denote
the distance between the $i$th well and the $j$th barrier
(on the right) by
$\Delta x_{i,j}=\widehat{x}_j-x_i$,
and their energy difference by
$\Delta E_{i,j}=\widehat{E}_j-E_i$.
We also define a formal (effective) rate constant
from the $i$th well over the $j$th barrier on the right as
\be
k_{i,j}=\omega_0 \, \alpha_i \, \widehat{\alpha}_j \,
 \e^{-\Delta E_{i,j}/\kT} .
\label{eq_k_eff}
\ee
Obviously $\Delta E_{i,j}$ and $k_{i,j}$ implicitly depend on $f$,
whereas $\Delta x_{i,j}$ are constants given the assumption of the previous paragraph.

%%%%%%%%%%%%%%%%%%%%%%%%%%%%%%%%%%%%%%%%%%%%%%%%%%%%%
\section{Stochastic kinetics of unbinding under external forces: standard description and corresponding approximations}\label{sec_std}
%%%%%%%%%%%%%%%%%%%%%%%%%%%%%%%%%%%%%%%%%%%%%%%%%%%%%%%%%%

We first recall the standard description of the "force spectrum", 
which relies on two major assumptions, namely the single escape rate (SER) and the deeply bound fundamental state (DBFS) approximations.  

%%%%%%
\subsection{Single escape rate approximation}
%%%%%%
If for any experimentally relevant load $f$ the
equilibration within the bound states is much faster than the escape to the
unbound state, the unbinding can be described by a single load
dependent escape rate $k(f)$. Following the calculation of Evans
\cite{Eva01}, if rebinding is negligible (which is the
case in most experimental situations), the probability $P(t)$ of
remaining in the bound state at time $t$ (the survival probability
of the bond) then decreases as
\begin{equation}
\frac{\dd P(t)}{\dd t} = -k(rt) P(t).
\end{equation}
The solution of this differential equation is
$P(t) = \exp[-\int_0^t k(rt') \, \dd t']$.
The probability density for unbinding between times
$t$ and $t+\Delta t$ is 
$p_t(t)=- \dd P(t)/\dd t= k(rt) P(t)$, from which, after changing the variable from $t$
to $f$, one gets the probability density
%\be
%p_f(f) = (1/r) k(f) \exp\left[-(1/r)\int_0^f \! k(f') \, \dd f' \right]
%\ee
for the distribution of the unbinding force: $p_f(f)=(1/r) k(f)P(f/r)$.
The typical unbinding force $f^*$ is defined as the peak of this
probability density:
$\dd p_f(f)/\dd f|_{f=f^*} =0$, which yields
the simple formula
\be
\left.\frac{\dd \tau(f)}{\dd f}\right|_{f=f^*}=-\frac{1}{r} ,
\label{eq_f}
\ee
where $\tau(f)\equiv 1/k(f)$ denotes the load dependent mean escape time. 
This formula gives the loading rate $r$ 
at which the typical unbinding force is $f^*$. For practical 
purposes it is often necessary to invert this relation to, e.g., predict the typical unbinding force for an experimentally imposed loading rate. 

To set a reference for further comparison,  we explicitely invert the above relation in case of a single barrier, i.e.\ when $\tau(f)=\tau(0)\exp(-f {\widehat x}_1)$, and obtain
\begin{equation}
 f^* = \frac{\kT}{{\widehat x}_1} \ln\left[\frac{r\tau(0){\widehat x}_1}{\kT}\right].
\end{equation}
 As mentioned in the introduction,
the escape over a single barrier results in a single straight line in the force spectrum. 
The experimental observation of a linear segment consequently
gives hints as to the structure of the energy landscape,
in particular the slope of the segment permits to deduce
a distance $\widehat{x}_1$ between the energy well and the barrier.

%%%%%%%%
\subsection{Deeply bound fundamental state approximation}
%%%%%%%%

Assuming further that the fundamental bound state is much deeper than the
intermediate ones: $E_i(f)-E_0(f)\gg\kT$ for any experimentally
relevant load $f$ (i.e. before unbinding has statistically almost certainly occurred, see Fig.~\ref{fig2} a.), 
Evans has shown that the mean escape time from the
fundamental bound state to the unbound state is well approximated by
\cite{Eva01,Eva02}
\be
\tau(f)
 =\sum_{j=1}^N
 \frac{1}{k_{0,j}(f)}
 =\sum_{j=1}^N
 \frac{\e^{-f\Delta x_{0,j}/\kT}}{k_{0,j}(0)}
\label{eq_evans}
\ee
This allows one to obtain an explicit $r$ vs.\ $f^*$  relationship
by plugging
Eq.\ (\ref{eq_evans}) into Eq.\ (\ref{eq_f}), 
which yields the compact formula
\begin{equation}
r=\left[
 \sum_{j=1}^N
 \frac{\Delta x_{0,j}}{\kT} \,
 \frac{\e^{-f^*\Delta x_{0,j}/\kT}}{k_{0,j}(0)}
\right]^{-1}.
\label{eq_fr_dbfs}
\end{equation}
This equation predicts a spectrum $f^*$ vs. $\log (r)$ consisting of a succession of
at most $N$ segments of increasing slopes, each of which yielding 
an information $\Delta x_{0,j}$ about an intermediate barrier.

%%%%%%%%%%%%%%%%%%%%%%%%%%%%%%%%%%%%%%%%%%%%%%%%%%%%%%%%%%%%%%
\section{Beyond the deeply bound fundamental state approximation}
%%%%%%%%%%%%%%%%%%%%%%%%%%%%%%%%%%%%%%%%%%%%%%%%%%%%%%%%%%%%%

In this section we relax the approximation made in the last subsection (IIIB),
generalize accordingly equations (7) and (9), and discuss the experimental implications of this generalization.
 
%%%%%%%%%%
\subsection{Refined theory}
%%%%%%%%%%
\begin{figure}%[!b]
\centerline{\includegraphics[width=\columnwidth]{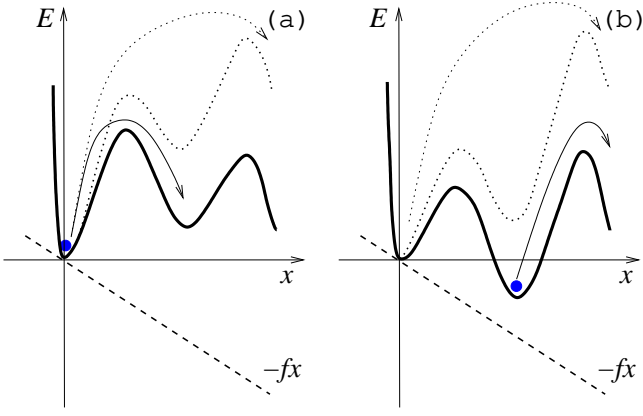}}
\caption{
Sketch of two energy landscapes with one intermediate well. Dotted drawings: no external force. When a constant force is applied energies are lowered by $fx$ (dashed lines), the resulting landscapes appear in solid lines. The dotted arrows indicate which pair of wells and barriers control the kinetics at zero load.  The solid arrows indicate the new limiting effective escape process at higher forces.
(a) The escape from the fundamental bound state remains the limiting process whatever the pulling force. 
(b) The escape form the intermediate bound state energy becomes the limiting process at high forces.
}\label{fig2}\end{figure}

In general, it is possible that for large enough forces one or more of the
intermediate bound states become deeper than the fundamental bound
state before unbinding has occured (see Fig.~\ref{fig2} b.). 
In such cases the above DBFS approximation breaks down. 
However, we show below that it is still possible to compute rather simply
the escape time $\tau$ from the
"bound state" to the "unbound state", provided
we maintain the assumption of 
a single escape rate $1/\tau(f)$. 

Let us put the system into its fundamental bound state, and let it
evolve according to the transition rates given in Eqs.\ (\ref{eq_k-})
and (\ref{eq_k+}). Whenever the system gets into the unbound state (by
making a transition over the outermost barrier) let us place it back into
the fundamental bound state. The stationary state of an ensemble of
such systems is characterized by a probability current, which is
constant everywhere and equal to $1/\tau$ by definition. 
To calculate $\tau$ we have to
solve the following system of equations:
\bea
P_i k^+_i - P_{i+1} k^-_{i+1} &=& 1/\tau \qquad 0\leq i \leq N-2 ,
 \label{eq_i}\\
P_{N-1} k^+_{N-1} &=& 1/\tau ,
 \label{eq_N-1}\\
\sum_{i=0}^{N-1} P_i &=& 1 ,
 \label{eq_norm}
\eea
where $P_i$ denotes the probability of being in the $i$th bound state
($0\leq i \leq N-1$). The first $N$ equations describe the probability
current over each of the $N$ barriers, and the last equation is just
the normalization condition. These $N+1$ linear equations uniquely
determine the $N+1$ variables ($P_i$ and $\tau$), and can be solved
easily in a recursive way. First, $P_{N-1}\tau$ can be expressed from
Eq.\ (\ref{eq_N-1}), and then $P_{N-2}\tau$, $...$, $P_0\tau$
recursively from Eq.\ (\ref{eq_i}) yielding
\bea
P_i \tau &=&
 \frac{1}{k^+_i} +
 \frac{k^-_{i+1}}{k^+_i k^+_{i+1}} + ... +
 \frac{k^-_{i+1} ... k^-_{N-1}}{k^+_i k^+_{i+1} ... k^+_{N-1}}
\nonumber\\
&=&
 %\frac{1}{k_{i,i+1}} + \frac{1}{k_{i,i+2}} + ... + \frac{1}{k_{i,N}}
%=
 \sum_{j=i+1}^N \frac{1}{k_{i,j}} \label{eq_Pitau},
\eea
where Eqs. (\ref{eq_k-}), (\ref{eq_k+}), and the definition 
(\ref{eq_k_eff}) have been
used. Note that because the $k_{i,j}$ are only formal definitions,
constructed as products
and ratios of the single-barrier rates (\ref{eq_k-}) and (\ref{eq_k+}),
they are meaningful even if
$\Delta E_{i,j}<0$.
From the normalization (\ref{eq_norm}) one can easily express $\tau$ as
\be
\tau
 =\sum_{i=0}^{N-1} P_i \tau
 =\sum_{i=0}^{N-1} \sum_{j=i+1}^N
 \frac{1}{k_{i,j}} .
\label{eq_tau_int}
\ee
The sum is dominated by the smallest effective rates, which are the
bottlenecks of the unbinding process. 
Consequently, this formula remains a good
approximation for $\tau$ even if some of the barriers disappear at big
loads, because the corresponding formal transition rates make negligible
contributions. By indicating the load force $f$ explicitly, we arrive
at
\begin{equation}
\tau(f)
 =\sum_{i=0}^{N-1} \sum_{j=i+1}^N
 \frac{1}{k_{i,j}(f)}
 =\sum_{i=0}^{N-1} \sum_{j=i+1}^N
 \frac{\e^{-f\Delta x_{i,j}/\kT}}{k_{i,j}(0)} .
\label{eq_tau}
\end{equation}
which generalizes (7).
An {\em analytic} formula can be given for the
$f^*$ vs.\ $r$ relationship by plugging
Eq.\ (\ref{eq_tau}) into Eq.\ (\ref{eq_f}) :
\begin{equation}
r=\left[
 \sum_{i=0}^{N-1} \sum_{j=i+1}^N
 \frac{\Delta x_{i,j}}{\kT} \,
 \frac{\e^{-f^*\Delta x_{i,j}/\kT}}{k_{i,j}(0)}
\right]^{-1},
\label{eq_fr}
\end{equation}
This generalization of equation (8) is one of the main results of this paper. 
%However, we keep in mind that the SER approximation may not hold since the equilibration between the bound states
%is not necessarily fast compared to the escape any more (see
%discussion later)
%(I would simply remove this remark, DB.).
Let us briefly comment on immediate features of this new formula.

First, Eq.~\eqref{eq_fr_dbfs} is easily recovered from (15) assuming a DBFS. Indeed, the assumption $E_i(f)\gg E_0(f)$ implies $k_{0,j}\gg k_{i,j}$, if $i>0$ [see Eq.~\eqref{eq_k_eff}] and therefore, the relation $P_i/P_0\ll 1$, if $i>0$ is deduced from Eq.~\eqref{eq_Pitau}. The probability to find the system in the fundamental bound state is close to $1$. So, the sum over $i$ in Eq.~\eqref{eq_tau} is dominated by the contributions of the effective escape rates from the $0$th well only. Finally, the sum over $i$ (labeling the intermediate states) is reduced to its sole first term too, and Eq.~\eqref{eq_fr} becomes identical to Eq~\eqref{eq_fr_dbfs}.

\begin{figure}[!t]
\centerline{\includegraphics[width=\columnwidth]{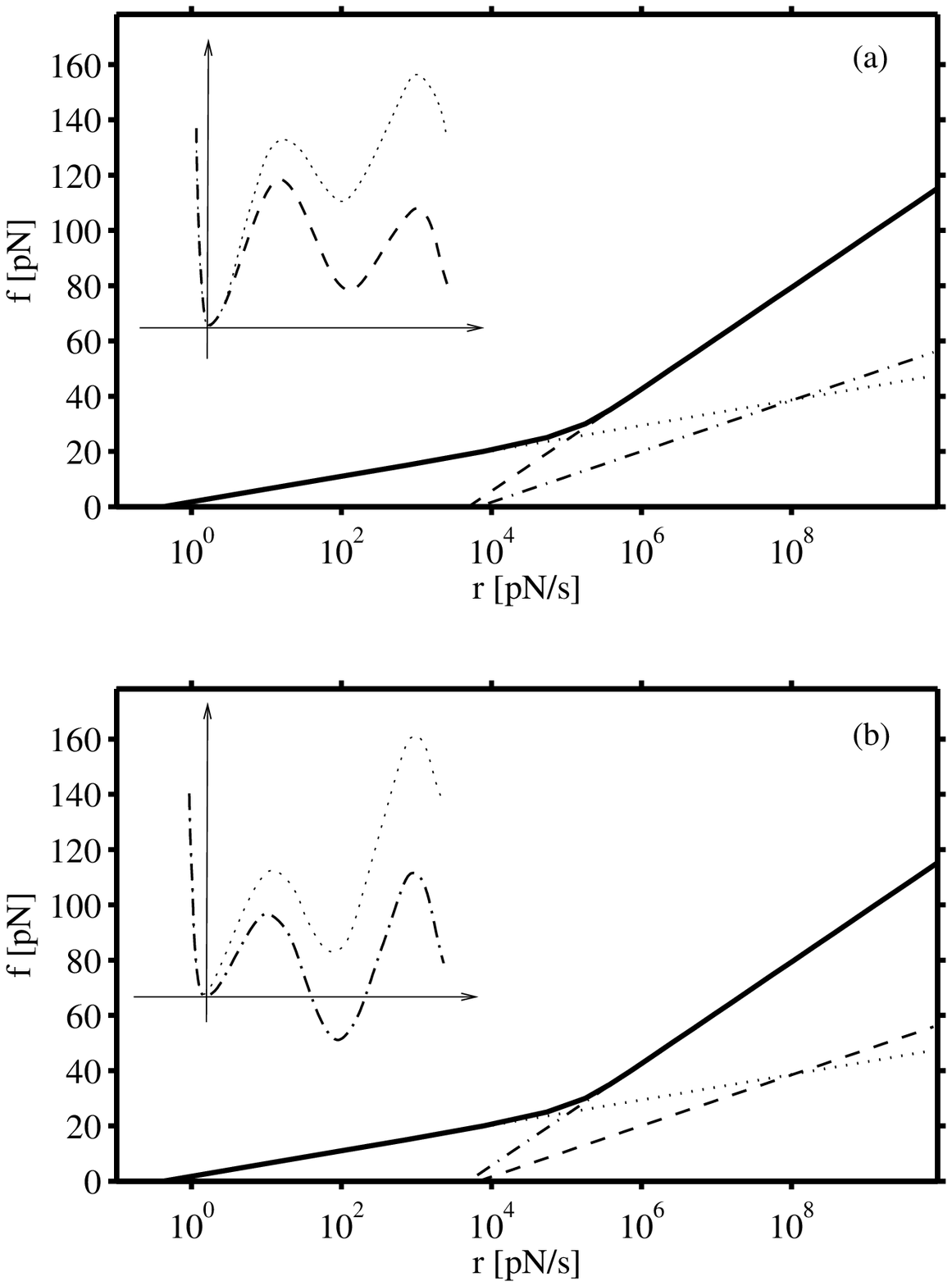}}
\caption{Two very similar force spectra  corresponding to different energy landscapes with one intermediate well.\\ 
Curves in solid lines: Force spectra plotted using Eq.~\eqref{eq_fr}. It closely follows the upper envelope of the straight lines corresponding to the transitions: $k_{0,2}$ (dotted line), $k_{0,1}$ (dashed line) and $k_{1,2}$ (dash-dotted line).\\ Insets: shape of the energy landscapes at low and high forces, each drawn with the same line style as the straight line  associated with the limiting transition.\\
Parameter values: 
(a) 
$(\widehat{x}_1,\widehat{E}_1)=(1\,{\rm nm},11\,\kT)$,
$(x_1,E_1)=(1.5\,{\rm nm},8 \,\kT)$, and
$(\widehat{x}_2,\widehat{E}_2)=(2\,{\rm nm},20\,\kT)$,
(b) 
$(\widehat{x}_1,\widehat{E}_1)=(0.5\,{\rm nm},12\,\kT)$,
$(x_1,E_1)=(1\,{\rm nm},9 \,\kT)$, and
$(\widehat{x}_2,\widehat{E}_2)=(2\,{\rm nm},20\,\kT)$.
}
\label{fig3}
\end{figure}

Second, each of the $N(N+1)/2$ terms of Eq.\ (\ref{eq_fr}) alone would yield
a straight line in the $f^*$ vs.\ $\log(r)$ plot. However, at any
loading rate the highest force value (the uppermost segment,
corresponding to the most
difficult transition) limits the unbinding process, therefore, the
$f^*(r)$ curve is expected to closely follow the upper envelope of these 
segments [see Fig.\ \ref{fig3}(a)]. Depending on the position of the
lines, this upper envelope can consist of up to $N(N+1)/2$ linear segments.

Third, this last point is clearly at odds with the
prediction within the DBFS approximation.
Indeed assuming a DBFS corresponds to 
forbidding the display in the force spectrum of 
the $N(N-1)/2$ segments 
corresponding to the probing of the escape from an intermediate bound state (see Fig. \ref{fig2}. b.).

%%%%%%%%%%%%%%%%%%%%%%%%%%%%%%%%% 
\subsection{Practical implications: Ambiguity in the determination of "structural" parameters}
%%%%%%%%%%%%%%%%%%%%%%%%%%%%%%%%%

We now insist on some practical implications 
of the above general description. 
We do not attempt a full inspection of all the possible dynamic responses of arbitrarily complex systems, but rather focus on two simple examples
in order to stress that the main features of the energy landscape can
in general {\em not} be unambiguously inferred from $[\log(r),f^*]$ plots. 
To emphasize the experimental relevance of this discussion, we use
for the parameters values comparable to those observed in experimental systems.
Specifically, we take the geometric factors
$\alpha_i$s and $\widehat{\alpha}_j$s to be all equal to $1$, $\omega_0=10^8$~s$^{-1}$ and $\kT=4\times10^{-21}$J.%
%, and express distances
%in nanometer, energies in $k_BT$, forces in pN and loading rates in pN/s.

\begin{figure}%[!t]
\centerline{\includegraphics[width=\columnwidth]{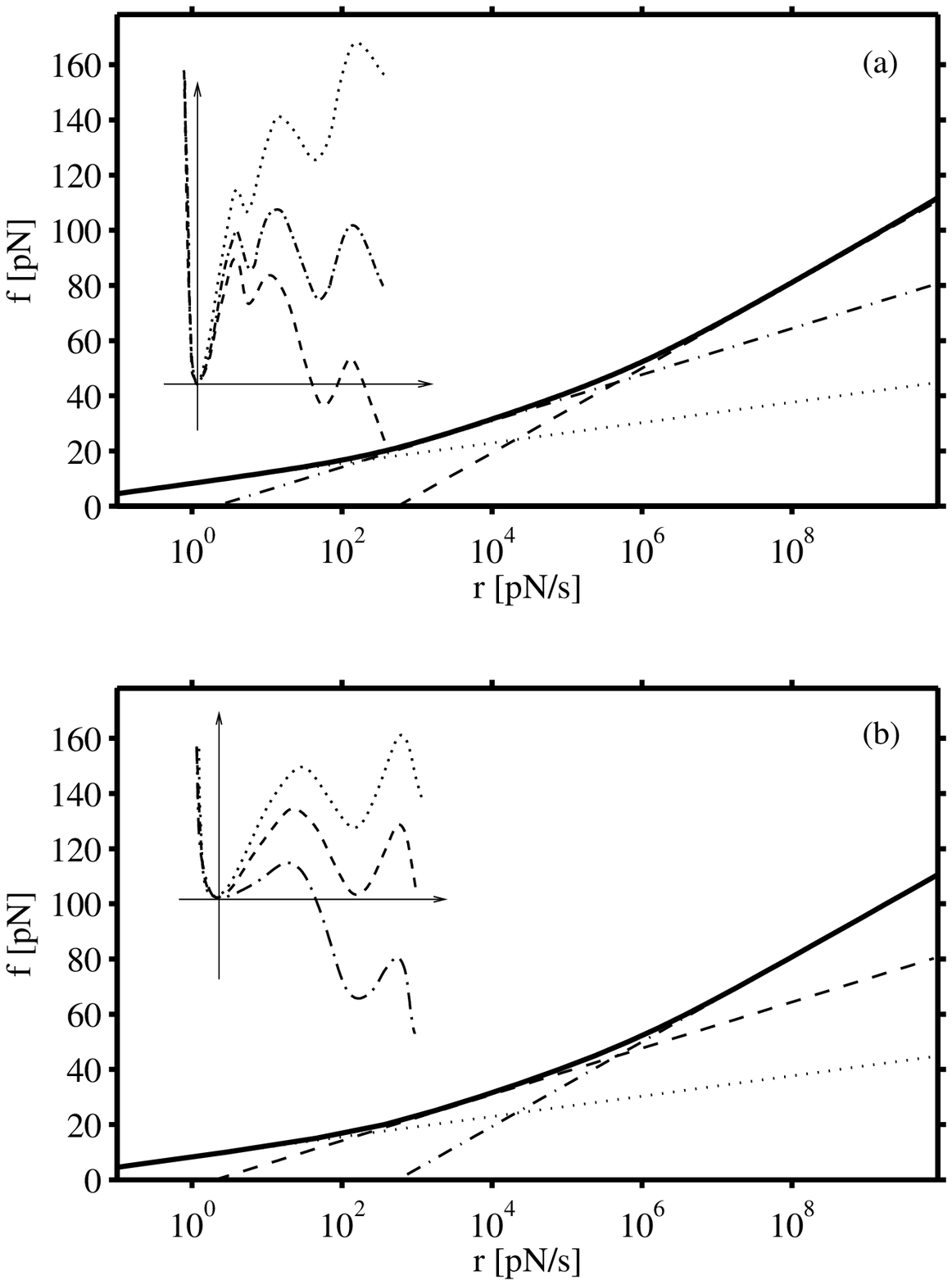}}
\caption{Two very similar force spectra corresponding to energy landscapes with different numbers of intermediate wells.\\ The rule of the line styles is the same as in Fig.~\ref{fig3}:
(a) dotted lines: $k_{0,3}$, dash-dotted lines: $k_{0,2}$, dashed lines: $k_{0,1}$. (b) dotted lines: $k_{0,2}$, dash-dotted lines: $k_{1,2}$, dashed lines: $k_{0,1}$.\\
Parameter values:
(a) 
$(\widehat{x}_1,\widehat{E}_1)=(0.6\,{\rm nm},14\,\kT)$,
$(x_1,E_1)=(0.7\,{\rm nm},12 \,\kT)$,
$(\widehat{x}_2,\widehat{E}_2)=(1.1\,{\rm nm},19\,\kT)$, 
$(x_2,E_2)=(2\,{\rm nm},16\,\kT)$ and
$(\widehat{x}_3,\widehat{E}_3)=(2.5\,{\rm nm},24\,\kT)$.
(b)
$(\widehat{x}_1,\widehat{E}_1)=(1.1\,{\rm nm},19\,\kT)$,
$(x_1,E_1)=(1.9\,{\rm nm},10 \,\kT)$, and
$(\widehat{x}_2,\widehat{E}_2)=(2.5\,{\rm nm},24\,\kT)$.
}
\label{fig4}
\end{figure}

%%%%%%%%%%%%%%%%%%
\subsubsection{Ambiguity in determining the barriers positions}
%%%%%%%%%%%%%%%%%%

Fig~\ref{fig3}\ (a) and Fig~\ref{fig3}\ (b) 
display two force spectra as obtained from Eq.~\eqref{eq_fr}. Both correspond to energy landscapes with two barriers.
Though the two $[\log(r),f^*]$ plots are almost identical 
they are related to very different 
sets of values for the energy levels and positions (along the pulling direction) 
of the wells and the barriers. 

Fig~\ref{fig3}\ (a) corresponds to the situation where the standard picture to account for the two segments is well suited~\cite{Eva}. At low force, the escape from the fundamental $0$th state over the outermost barrier is the limiting process. The slope of the first segment is proportional to $k_BT/\widehat{x}_2$. For the highest forces (above $\sim 30{\rm pN}$) the energy of the external barrier is reduced below $\widehat{E}_1$ and the deepest bound state remains located at $x_0=0$. The process that mostly impedes the unbinding is the overcome of the innermost barrier $\widehat{E}_1$ with a rate $k_{0,1}$. The slope of the curve is now larger and proportional to $k_BT/\widehat{x}_1$.

Fig. \ref{fig3}\ (b) corresponds to an energy landscape for which the above explanation is inappropriate. At low force the unbinding kinetic is controlled by the escape from the fundamental state over the outermost barrier again. But, for pulling forces larger than $\sim 30{\rm pN}$ this outer barrier remains the highest [see inset in Fig~\ref{fig3}.(b)]. However the slope of the spectrum increases as in the (a) case. The reason is that the deepest (and most occupied) bound state is now located at $x=x_1$ and the presence of the second segment actually witnesses the escape from this intermediate state to the unbound state with a rate $k_{1,2}$. The value of the second slope scales therefore with $k_BT/(\widehat{x}_2-x_1)$. Since the escape rate $k_{1,2}$ in the (b) case is equal to the escape rate $k_{0,1}$ in the (a) case, the two spectra in Fig.~\ref{fig3} turn out to be indistinguishable and cannot be {\it a priori} associated with one of the two possible landscapes. 

%%%%%%%%%
\subsubsection{Ambiguity in determining the number of barriers}
%%%%%%%%%%

After having shown with the simple example above that ambiguity 
can exist in determining distances from dynamic force spectra, we show here
that even more strikingly it is impossible in general to assess the number of wells and barriers. Again we use a simple example to do so.

Fig. \ref{fig4} displays two force spectra obtained using Eq.~\eqref{eq_fr}. They are both well approximated by a succession of three segments with increasing slopes. Again, the two $[\log(r),f^*]$ curves are very similar although they are constructed from landscapes that do not even comprise the same number of peaks and wells.

In Fig.~\ref{fig4}\ (a) the three segments describe 
the escape from the same fundamental state over the three distinct energy barriers. The larger the pulling force the closer the limiting barrier to the fundamental state [see inset in Fig.~\ref{fig4}\ (a)].

In Fig.~\ref{fig4}\ (b), the landscape consists of only two barriers. 
However, the force spectrum reveals that three different escape processes can limit the unbinding kinetic. At low forces ($f\lesssim50 {\rm pN}$) the two observed linear segments results from the escape form the fundamental state over the two peaks at $\widehat{x}_1$ and $\widehat{x}_2$ respectively. Conversely, at high forces it is the escape from the deeply lowered intermediate state over the outer barrier that determines the escape rate (see inset in Fig. \ref{fig4}\ (b), drawing with dash-dot line). With the chosen parameters the effective rate $k_{0,3}$, $k_{0,2}$ and $k_{0,1}$ in Fig. 4(a) case correspond respectively to $k_{0,2}$, $k_{0,1}$ and $k_{1,2}$ in Fig. 4(b) case. Thus the two plots are indistinguishable and cannot be used to predict the number of barriers along the 1D escape path.

In conclusion of this subsection, we suggest great care in inferring features
of the underlying energy landscape from dynamic force spectroscopy experiment.
Our generalized equation may 
be helpful in dealing with the corresponding ambiguity
as it allows (with some work) to generate various landscapes
that can account for the observed data, whereas 
Eq.~\eqref{eq_fr_dbfs} can only yield a single set of parameters (
e.g. those used for the plots in Fig.~\ref{fig3} (a) and Fig.~\ref{fig4} (a)).

%In summary the two major implications of this first step toward a generic description of the unbinding %kinetic are: 
%(i) It is impossible to extract the absolute position of the barriers along the pulling direction from a %force spectrum composed of more than one segment without complementary knowledges on the topography of the %energy landscape. Only the separations $\Delta x_{i,j}$ can be inferred unambiguously.
%(ii) The number of distinct relevant barriers along the pulling direction do not necessarily corresponds to %the number of linear segments observed in the $[\log(r),f^*]$ plots but only to the number of different %effective escape processes that mostly limit the unbinding kinetic.

%%%%%%%%%%%%%%%%%%%%%%%%%%%%%%%%%%%%%%%%%%%%
\section{Beyond the single escape rate approximation: Multimodal unbinding force distributions}
%%%%%%%%%%%%%%%%%%%%%%%%%%%%%%%%%%%%%%%%%%%%%%%
\begin{figure}[!t]
\centerline{\includegraphics[width=\columnwidth]{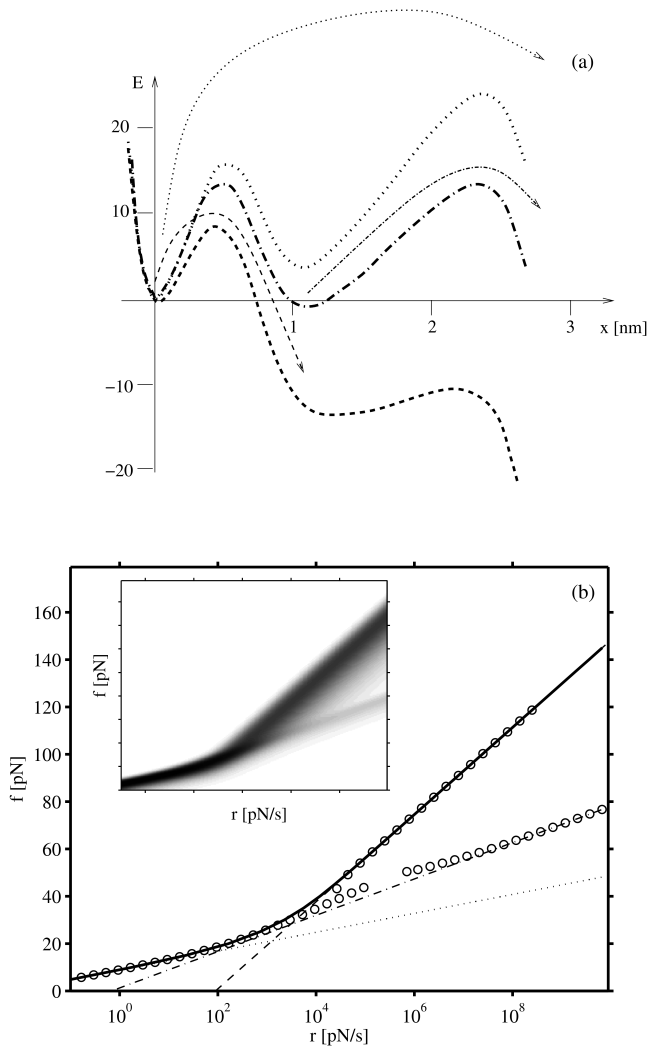}}%{fig5Pab22}}
\caption{A scenario that yields a multimodal unbinding force distribution.(a) Three snapshots of the energy landscape for the pulling forces: $f=0\,{\rm pN}$ (dotted line), $f=18\,{\rm pN}$ (dahs-dotted line) and $f=60\, {\rm pN}$ (dashed line). The three arrows indicate the corresponding most difficult transitions. Parameters values: $(\widehat{x}_1,\widehat{E}_1)=(0.5\,{\rm nm},16\,\kT)$,
$(x_1,E_1)=                        (1.1\,{\rm nm},4 \,\kT)$, and
$(\widehat{x}_2,\widehat{E}_2)=(2.3\,{\rm nm},24\,\kT)$.
(b) Force spectrum (full line) associated with the landscape described in (a) plotted using Eq.~\eqref{eq_fr}.
The full probability density $p_f(f)$ for unbinding at force $f$ is plotted in gray scale in the inset.
$p_f(f)$ has been obtained following the procedure described in~\cite{Bar02}. The circles in the main plot represent the local maxima of the unbinding force distribution $p_f(f)$ 
}
\label{fig5}
\end{figure}

Up to this point we have been considering a generalized theory in which
the deeply bound fundamental state (DBFS) approximation is dropped, but
the unbinding is still approximated as a simple first-order escape
process. Indeed, the validity of Eq.\ (\ref{eq_fr}) relies on the
assumption that at any moment the distribution of the populations of
the bound states can be well approximated by the distribution
corresponding to a homogeneous stationary current.

This is, however, not always the case. As we stated earlier, the sum of
the
$1/k_{i,j}(f)$ terms in Eq.\ (\ref{eq_tau}) is dominated by the
smallest effective rate constant
$k_{i',j'}(f)$
corresponding to the slowest effective transition. A consequence of
this is that all the bound states located to the left of barrier $j'$
are close to equilibrium (because of the slow outflow over barrier $j'$),
and the population of any state located to the right is negligible
(because they practically belong to the unbound state). Now, If the
slowest transition rate changes from
$k_{i',j'}(f)$ to $k_{i'',j''}(f)$ as the loading force $f$ is
increased, and
if $j''<j'$, then a considerable population might remain in the
intermediate bound states between the new and the old limiting
barriers, $j''$ and $j'$ respectively. This residual population is
incompatible with the new stationary current dominated by
$k_{i'',j''}(f)$, and must escape in a different
way, yielding a secondary maximum of the unbinding force distribution
(see Fig.\ \ref{fig5}b).

The escape of the majority of the population (located to the left of
the new limiting barrier $j''$) can still be characterized by Eq.\
(\ref{eq_tau}) of our generalized theory. On the other hand, we have to
slightly modify this formula to describe the escape of the residual
population (trapped between the new and old limiting barriers). Since
$j''$ is the limiting barrier now, almost the entire residual
population can escape without ever jumping backward over barrier $j''$.
Therefore, for the residual population we can consider barrier $j''$ as
a reflecting boundary, and describe the escape by our general theory in
this modified potential. Eq.\ (\ref{eq_tau}), e.g., changes accordingly:
\begin{equation}
\tau(f)
 =\sum_{i=j''}^{N-1} \sum_{j=i+1}^N
 \frac{1}{k_{i,j}(f)}
 =\sum_{i=j''}^{N-1} \sum_{j=i+1}^N
 \frac{\e^{-f\Delta x_{i,j}/\kT}}{k_{i,j}(0)} .
\label{eq_tau_bis}
\end{equation}

Consequently, the absolute maximum of the unbinding force distribution
always follows the upper envelope of the $N(N+1)/2$ lines, however,
some secondary maxima might also appear at lower forces, which follow
the upper envelope of only a subset of the lines [comprising
$(N-j'')(N-j''+1)/2$ elements]. Such secondary maxima of a multimodal
force distribution give important information on the internal structure
of the energy landscape of the unbinding path, and makes the
determination of the number and positions of the energy wells and
barriers less ambiguous. 
It is actually a nice achievement of our
generalized theory to be able to make sense of the segments 
of secondary maxima in a unique frame for fitting parameters 
(see e.g. Fig. 5, where the segment corresponding to the secondary maximum
corresponds to the transition from the well 1 over the barrier 2, 
a step neglected in the DBFS approximation).
The possibility of a bimodal distribution for
the case of a two-state system 
has already been reported by Strunz et al.
\cite{Str00}, and our description systematizes and generalizes their findings.

To provide a simple illustration for the somewhat formal discussion
above,
we also focus on a system consisting of two
bound states, as depicted in Fig.\ \ref{fig5} (a).
Increasing the force, the limiting transition rate changes from
$k_{0,2}$ to $k_{1,2}$ and then to $k_{0,1}$. In the range of the
loading rate $r$ between about $10^4$ and $10^5$~pN/s the intermediate
bound state (1) has enough time to accumulate a large population, which is
then flushed by the $k_{1,2}$ transition before the transition
$k_{0,1}$ flushes the rest from the fundamental bound state (0). In the
range above $10^6$~pN/s the intermediate bound state (1) cannot
accumulate much of the population, but it still possesses a small
fraction of the initial equilibrium distribution, which
is again flushed by the $k_{1,2}$ transition first.

Very recently the group of 
Evans actually reported the experimental
occurrence of a bimodal force distribution
\cite{Eva02}.
The corresponding experiment consisted in pulling on "diC14 PE" lipids 
from a bilayer made of "C18:0/1 PC" lipids. 
With the help of our generalized theory,
Evans and Williams were able to fit their data and
interpret the results in terms of an energy landscape with 
two barriers [personal communication,
see Ref.\ [3] in Ref.\ \cite{Eva02}].

\section{Conclusion}
In this paper, we have revisited the standard theory used to account for the dynamic response of molecular stickers. Our refined description, valid for an arbitrarily complex one-dimensional energy landscape, has allowed us to highlight several practical consequences of the diversity of the possible unbinding scenarios. For example several markedly different energy landscapes can yield the same rupture force distribution. To resolve this ambiguity other experimental technics, e.g. flow chamber experiments~\cite{Pie02}, are then required. 
We have also identified the physical origin of multimodal unbinding force distributions and shown how their analysis provides informations on the unbinding pathways. 
%We hope that this theoritical analysis will prove usefull in the planning of future dynamic %force spectroscopy experiments.

%(i) It is impossible to extract the absolute position of the barriers along the pulling direction from a force spectrum composed of more than one segment without complementary knowledges on the topography of the energy landscape. Only the separations $\Delta x_{i,j}$ can be inferred unambiguously. Hence, other investigation method need to be  

%(ii) The number of distinct relevant barriers along the pulling direction do not necessarily corresponds to the number of linear segments observed in the $[\log(r),f^*]$ plots but only to the number of different effective escape processes that mostly limit the unbinding kinetic.
%(iii)multimodal

\vspace{-5pt}							%%% DELETE %%%

%\bibliography{p}

\begin{thebibliography}{}
\bibitem{Str03}
T. R. Strick, M.-N. Dessinges, G. Charvin, N. H. Dekker, J.-F. Allemand, D. Bensimon and V. Croquette
%Streching of macromolecules and proteins
Rep. Prog. Phys. {\bf 66}, 1 (2003).
%
\bibitem{titin}
 M. Rief, M. Gautel, F. Oesterhelt, J. M. Fernandez, and H. E. Gaub,
 % Reversible Unfolding of Individual Titin Immunoglobulin Domains by AFM
 Science {\bf 276}, 1109 (1997);
%
 M. S. Kellermayer, S. B. Smith, H. L. Granzier, and C. Bustamante,
 % Folding-Unfolding Transitions in Single Titin Molecules
 % Characterized with Laser Tweezers
 {\em ibid.}\ {\bf 276}, 1112 (1997).

\bibitem{Poi01}
 M. G. Poirier, A. Nemani, P. Gupta, S. Eroglu, and J. F. Marko,
 % Probing Chromosome Structure with Dynamic Force Relaxation
 Phys.\ Rev.\ Lett.\ {\bf 86}, 360 (2001).

\bibitem{Sim99}
 D. A. Simons, M. Strigl, M. Hohenadl, and R. Merkel,
 %Statistical breakage of single proteine A-IgG Bonds reveals
 %Crossover from spontaneous to Force-Induced Bond Dissociation
 Phys.\ Rev.\ Lett.\ {\bf 83}, 652 (1999).

\bibitem{Nis00}
 T. Nishizaka, R. Seo, H. Tadakuma, K. Kinosita, and S. Ishiwata,
 %Characterization of Single Actomyosin Rigor Bonds: Load Dependence of Lifetime
 %and Mechanical Properties
 Biophys.\ J.\ {\bf 79}, 962 (2000).

\bibitem{Pie96}
 A. Pierre, A. M. Benoliel, P. Bongrand, and P. A. van der Merwe,
 %Determination of the lifetime and force dependence of interactions of single bonds
 %between surface-attached CD2 and CD48 adhesion molecules
 Proc.\ Natl.\ Acad.\ Sci.\ USA {\bf 93}, 15114 (1996).

\bibitem{Eva}
 E. Evans and K. Ritchie,
 % Dynamic strength of molecular adhesion bonds
 Biophys.\ J.\ {\bf 72}, 1541 (1997);
%
 R. Merkel, P. Nassoy, A. Leung, K. Ritchie, and E. Evans,
 % Energy landscapes of receptor-ligand bonds explored with dynamic
 % force spectroscopy
 Nature {\bf 397}, 50 (1999).

\bibitem{Hum01}
 G. Hummer and A. Szabo,
 %Free energy reconstruction from nonequilibrium single-molecule pulling experiment
 Proc.\ Natl.\ Acad.\ Sci.\ USA {\bf 98}, 3658 (2001).

\bibitem{Hey00}
H. Grubm\"uller, B. Heymann and P. Tavan,
% Ligand binding: molecular mechanics calculation of the streptavidin-biotin rupture force.
Science,\ {\bf 271}, 997 (1996).

\bibitem{Eva01}
 E. Evans,
 % Energy landscapes of biomolecular adhesion and receptor anchoring at
 % interfaces explored with dynamic force spectroscopy
 Faraday Discuss.\ {\bf 111}, 1 (1998);
%
 E. Evans,
 % Probing the relation between force lifetime and chemistry in single
 % molecular bonds
 Annu.\ Rev.\ Biophys.\ Biomol.\ Struct.\ {\bf 30}, 105 (2001).

\bibitem{Sei02}
U. Seifert, 
Europhys. Lett.\ {\bf 58}, 792 (2002).

\bibitem{Bar02}
D. Bartolo, I. Der\'enyi and A. Ajdari,
%Dynamic response of adhesion complexes: Beyond the single path picture
Phys. Rev. E. {\bf 65}, 051910-1 (2002).

\bibitem{Str00}
 T. Strunz, K. Oroszlan, I. Schumakovitch, H.-J. G\"untherodt, and M. Hegner,
 % Model Energy Landscapes and the Force-Induced Dissociation of
 % Ligand-Receptor Bonds
 Biophys.\ J.\ {\bf 79}, 1206 (2000).

\bibitem{Eva02}
 E. Evans and P. Williams,
 % Dynamic Force Spectroscopy,
 in {\it Physics of bio-molecules and cells},
 edited by H. Flyvbjerg {\it et al.}
 (Springer, Berlin, 2002), pp. 145--204.

\bibitem{Pie02}
A. Pierres, D. Touchard, A.-M. Benoliel, and P. Bongrand,
%Dissecting steptavidin-biotin interaction with laminar flow chamber
Biophys.\ J.\ {\bf 82}, 3214 (2002).
\end{thebibliography}

\end{document}